\documentclass[twocolumn, preprintnumbers,amsmath,amssymb]{revtex4}

\usepackage{amsmath,amssymb,amsfonts} 	
\usepackage[dvips]{graphicx}
\usepackage[latin1]{inputenc}
\usepackage{subfigure}
\begin{document}

\title{Quantum corrected Langevin dynamics for adsorbates on metal surfaces interacting with hot electrons}

\author{Thomas Olsen}
\email{tolsen@fysik.dtu.dk}
\author{Jakob Schi{\o}tz}

\affiliation{Danish National Research Foundation's Center for Individual
Nanoparticle Functionality (CINF),
	Department of Physics,
	Technical University of Denmark,
	DK--2800 Kongens Lyngby,
	Denmark}

\date{\today}

\begin{abstract}
We investigate the importance of including quantized initial conditions in Langevin dynamics for adsorbates interacting with a thermal reservoir of electrons. For quadratic potentials the time evolution is exactly described by a classical Langevin equation and it is shown how to rigorously obtain quantum mechanical probabilities from the classical phase space distributions resulting from the dynamics. At short time scales, classical and quasiclassical initial conditions lead to wrong results and only correctly quantized initial conditions give a close agreement with an inherently quantum mechanical master equation approach. With CO on Cu(100) as an example, we demonstrate the effect for a system with ab initio frictional tensor and potential energy surfaces and show that quantizing the initial conditions can have a large impact on both the desorption probability and the distribution of molecular vibrational states.
\end{abstract}

\maketitle

\section{Introduction}
Femtosecond lasers has proven a most valuable tool in the study of excited metallic electrons and their interactions with surface adsorbates. In Ref. \onlinecite{prybyla1} it was shown that a femtosecond laser pulse could be used to desorb NO from Pd(111) and a mechanism involving multiple electronic excitations of the adsorbate was identified.\cite{budde, misewich1} Since then, it has been demonstrated that several other surface reactions can be induced by femtosecond laser pulses\cite{prybyla2,kao1,kao2,struck,ho,stepan,bonn1} and the mechanism is usually attributed to a direct interaction of excited (hot) metallic electrons interacting with adsorbate resonant states, although substrate heating may also contribute to reaction rates.\cite{wagner08}

A variety of theoretical models have been proposed to describe the interaction and resulting transfer of energy from hot electrons to adsorbates, but a common conceptual picture is can be given in terms of Born-Oppenheimer potential energy surfaces. It is then assumed that the adsorbate propagation is governed by a potential energy surface $V_0$ when the adsorbate is in its electronic ground state. If the adsorbate has a resonance (possibly partly occupied in the ground state), a hot metallic electron can transiently occupy the resonant state and the adsorbate dynamics will then be governed by a different potential energy surface $V_1$. Hot electrons can thus transfer energy to the adsorbate by inducing jumps between the two potential energy surfaces.\cite{misewich1} Although the lifetime of the excited electronic state on the adsorbate may be very short ($<1\;fs$), several such events can eventually transfer enough energy for the adsorbate to overcome a reaction barrier.
 
The probability that a hot electron scatters inelastically on the adsorbate and transfers a given amount of energy can be calculated in a local polaron model\cite{wingreen,gadzuk91,olsen1,olsen2} and may be generalized to reactions resulting from multiple electronic excitations.\cite{olsen3} However, since we are usually only interested in the adsorbate dynamics, it is often more convenient to apply open system density matrix theory. In this formalism, it is assumed that the femtosecond laser pulse gives rise to a hot thermalized distribution of electrons with a time dependent electronic temperature $T_e$. The time dependent density matrix of the full interacting system is then constructed and the electronic states are traced out resulting in a reduced density matrix with a diagonal that gives the probabilities that the adsorbate is in a particular state. Based on the Feynman-Vernon theory of influence functionals,\cite{feynman-vernon,caldeira,schmid} it is possible to calculate the reduced density matrix of a Newns-Anderson type Hamiltonian in either a coordinate basis\cite{brandbyge} leading to Langevin dynamics or in a basis of vibrational eigenstates\cite{gao97} leading to a master equation for the vibrational eigenstates. For a harmonic potential with frequency $\omega_0$, the master equation reduces to a Fokker-Planck equation in the classical limit of $k_BT_e\gg\hbar\omega_0$ and desorption probabilities can be obtained from an Arrhenius type expression.\cite{misewich2} However, as shown explicitly in Refs. \onlinecite{gao95} and \onlinecite{gao97}, the Fokker-Planck equation fails dramatically when the classical condition above is not satisfied and in general a quantum mechanical treatment of the adsorbate is needed. On the other hand, the coordinate representation of the reduced density matrix results in semi-classical dynamics for the adsorbate coordinates and the quantum nature of the problem only enters through the initial state. 

Langevin dynamics have been applied with reasonable success to problems involving hot electron induced surface reactions\cite{tully,luntz2} and to elucidate the role of non-adiabatic effects in general.\cite{luntz1,trail} However, the initial quantum state is usually neglected or treated quasiclassically. The purpose of the present work is to investigate the role of quantum mechanical boundary conditions and compare the results to those obtained with classical and quasiclassical initial states where only the zero point energy is included. In particular, we will focus on the harmonic oscillator since, when the initial state is included correctly, Langevin dynamics with a quadratic potential is exact to second order in perturbation theory and we can thus compare with a quantum mechanical master equation.

The paper is organized as follows. In section \ref{Theory} we introduce the model Hamiltonian which constitutes the foundation of the calculations. The time dependent density matrix of the harmonic oscillator is then reviewed and is shown to give rise to classical dynamics with quantum corrections entering only through the initial state which must be included by a phase space sampling procedure. Generalizing this approach to our model Hamiltonian results in Langevin dynamics with explicit expressions for the electronic friction tensor and correlations between fluctuating forces. In section \ref{model} we start by analyzing the harmonic oscillator and show how to obtain the quantum mechanical probabilities from the classical phase space distribution resulting from a Langevin equation approach. It is demonstrated that, when the initial conditions is correctly taken into account, the results show excellent agreement with the master equation approach. The comparison is then repeated for the Morse potential where the Langevin dynamics does not provide an exact description of the quantum dynamics, but which has the advantage of having a well defined desorption energy. In section \ref{ab_initio} we consider the example of hot electron induced desorption of CO from Cu(100) using ab initio potential energy surfaces, and perform Langevin dynamics with classical, quasiclassical, and quantum mechanical initial conditions. In appendix \ref{path}, it is  show how classical dynamics and the initial Wigner phase space distribution emerges from a path integral representation of the time dependent reduced density matrix in a quadratic potential.

\section{Theory}\label{Theory}
\subsection{Hamiltonian}
The Langevin dynamics with local electronic friction can be derived from a Newns-Anderson\cite{newnsanderson1,newnsanderson2} type Hamiltonian where a single adsorbate resonant state $|a\rangle$ is coupled to the adsorbate degrees of freedom $x_i$.\cite{brandbyge} The resonant state is usually chosen as an eigenstate of the adsorbate far from the surface. Close to the surface, $|a\rangle$ becomes hybridized with metallic states and acquires a finite lifetime. In the electronic ground state, the resonant state has a partial (or zero) occupation and the adsorbate propagation is governed by a ground state Born-Oppenheimer potential energy surface $V_0(x_i)$ with a local minimum at $x_i^0$. However, the presence of hot metallic electrons may give rise to a transient full occupation of the resonant state and the adsorbate propagation will then be governed by the potential energy surface $V_1(x_i)$. Even though the resonant state is short lived, a transient occupation will perturb the system and may result in a transfer of energy to the adsorbate\cite{olsen1}. The Hamiltonian describing the system can then be modelled by\cite{wingreen, gadzuk91, olsen1}
\begin{align}\label{H}
&H=H_{el}+H_0+H_I,\\
&H_{el}=\varepsilon_0c_a^{\dag}c_a + \sum_k\epsilon_kc_k^{\dag}c_k+\sum_kV_{ak}^0c_a^{\dag}c_k+h.c.\notag\\
&H_0=\sum_i\frac{p_i^2}{2M_i}+V_0(x_i)\notag\\
&H_I=\big(\varepsilon_a(x_i)-\varepsilon_0\big)c_a^{\dag}c_a+\sum_k\big(V_{ak}(x_i)-V_{ak}^0\big)c_a^{\dag}c_k+h.c.\notag\\
&\qquad\qquad\qquad\varepsilon_a(x_i)=V_1(x_i)-V_0(x_i)\notag
\end{align}
where $c_a^\dag$ and $c_k^\dag$ are creation operators for the resonant state $|a\rangle$ and metallic states $|k\rangle$ respectively and $\varepsilon_0=\varepsilon_a(x_i^{0})$, $V_{ak}^0=V_{ak}(x_i^0)$. Conceptually, the Hamiltonian describes an adsorbate with dynamics governed by $V_0(x_i)$ in the electronic ground state and $V_1(x_i)$ when the resonant state is occupied, and the reservoir of metallic electrons can exchange energy with the adsorbate via the resonant state. The hybridization depends on the position of the adsorbate through $V_{ak}(x_i)$ which become zero when the adsorbate is far from the surface. It should be noted that if $V_{ak}$ are constant and the ground and excited state potentials are quadratic with displaced minima, one obtains $H_I=-c_a^\dag c_a\sum_if_ix_i$. The coupling constants are then given by $f_i=m_i\omega_i^2\tilde{x}_i$ where $\tilde{x}_i$ is the shift in the minimum of the excited state potential with respect to the ground state minimum. 

We will impose the wide band limit in which the metallic band coupled to the adsorbate is assumed to be much wider than the resonance width. For a fixed position of the adsorbate, the density of states projected onto the resonance is then a Lorentzian:
\begin{align}
\rho_a(\varepsilon)=\frac{1}{\pi}\frac{\Gamma/2}{(\varepsilon-\varepsilon_a)^2+(\Gamma/2)^2},
\end{align}
with the full width at half maximum given by
\begin{align}
\Gamma=2\pi\sum_k|V_{ak}|^2\delta(\varepsilon_a-\epsilon_k).
\end{align}
In these expressions both $V_{ak}$ and $\varepsilon_a$ and therefore $\rho_a$ and $\Gamma$ depend parametrically on the instantaneous position of the adsorbate.


\subsection{The density matrix}
The advantage of the density matrix formalism is two-fold. First of all, for complicated systems one may trace out all irrelevant degrees of freedom from the density matrix and the resulting 'reduced' density matrix then describes a system which can exchange energy with the environment. Second, the density matrix formalism allows one to treat a statistical ensemble of states in a natural way. In the case of an adsorbate interacting with electrons in a metal, as described by the Hamiltonian \eqref{H}, the full density matrix can be reduced by tracing out the electronic degrees of freedom and the diagonal elements of the resulting reduced density matrix then gives the probabilities of finding the adsorbate in a particular state as a function of time.

The time dependent density matrix is
\begin{align}\label{rho_t}
\rho(t)=e^{-iHt/\hbar}\rho_0e^{iHt/\hbar},
\end{align}
where $\rho_0$ is the density matrix at $t=0$. As always it is instructive to consider a harmonic oscillator and we thus start by considering $H_0$ of Eq. \eqref{H} with a single degree of freedom and a quadratic potential. In the coordinate representation the density matrix can then be written
\begin{align}\label{rho_harmonic}
\rho(x,y;t)=\langle x|\rho(t)|y\rangle=&\int dx_0dy_0\langle x_0|\rho_0|y_0\rangle\\
&\times\langle x|e^{-iH_0t/\hbar}|x_0\rangle\langle y_0|e^{iH_0t/\hbar}|y\rangle.\notag
\end{align}
We see that the density matrix involves two propagators and the integrand can be viewed as a particle first being propagated forward in time from $x_0$ to $x$ and then backward in time from $y$ to $y_0$. The propagator of the harmonic potential is well known\cite{sakurai} and the result for the diagonal elements is
\begin{align}\label{density_harmonic}
\rho(u;t)=&\int du_0dp_0\mathcal{P}(u_0,p_0)\notag\\
&\times\delta\Big(u(t)-[u_0\cos\omega t+p_0\frac{\sin\omega t}{m\omega}]\Big)
\end{align}
where
\begin{align}\label{wigner}
\mathcal{P}(x,p)=\frac{1}{2\pi\hbar}\int dy\langle x-y/2|\rho_0|x+y/2\rangle e^{ipy/\hbar}
\end{align}
is the Wigner distribution of an initial state described by the density matrix $\rho_0$ and $u=x=y$. The Wigner distribution is often referred to as a quasi probability distribution and can be interpreted as the quantum mechanical probability of finding a particle in the small phase space area $dxdp$.\cite{hillery} This means that the expression \eqref{density_harmonic} can be thought of as a sum over all initial phase space configurations weighted by their probabilities and subject to the constraint dictated by the delta function. However, the constraint is equivalent to the Newtonian equations of motion and we can thus regard the time evolution as purely classical. In particular, given an initial state we could calculate $\rho(u;t)$ by sampling all phase space and adding $\mathcal{P}(u_0,p_0)$ if $u_0$ and $p_0$ is classically connected to $u(t)$. Furthermore, since each such classical trajectory will result in a well defined momentum at time $t$ we interpret the probability of being at a given phase space point $u(t),p(t)$ as being equal to $\mathcal{P}(u_0,p_0)$ where $(u_0,p_0)$ is the unique point which is classically connected to $(u(t),p(t))$. The quantum nature of the particle propagating in a harmonic oscillator potential thus solely enters through the initial state specified by $\rho_0$. This is of course closely related to the well known fact that for a harmonic potential, the time evolution of the Wigner distribution is equal to the time evolution of a classical phase space distribution.\cite{hillery}

The Langevin equations emerge when the electronic degrees of freedom is traced out from the time dependent density matrix corresponding to the full Hamiltonian \eqref{H}. With a quadratic potential the result is very similar to \eqref{density_harmonic} the only difference being that the coupling to a thermal reservoir of electrons introduces a broadening in the delta function. Thus the time evolution can be thought of as classical with fluctuations that has a magnitude determined by the broadening. It has previously been shown that these fluctuations can be handled in a  statistical sense\cite{caldeira, schmid} and the full dynamics can be written in terms of classical equations of motion with a stochastic force $\xi_i(t)$. The stochastic force is specified by its statistical properties which is related to the broadening of the delta function. The result is the Langevin equation
\begin{align}\label{Langevin}
M_i\ddot{u}_i+\frac{d}{du_i}V_0(u)+\sum_j\eta_{ij}(u)\dot{u}_j=\xi_i(t)
\end{align}
where the local temperature dependent friction tensor is given by
\begin{align}\label{friction}
\eta_{ij}(u)=\frac{-\hbar}{\pi}\int_{-\infty}^\infty& d\varepsilon\bigg(\frac{\Gamma(u)/2}{(\varepsilon-\varepsilon_a(u))^2+(\Gamma(u)/2)^2}\bigg)^2\notag\\
&\times f_i(\varepsilon;u)f_j(\varepsilon;u)\frac{dn_F(T;\varepsilon)}{d\varepsilon}
\end{align}
with
\begin{align}\label{fric_force}
f_i(\varepsilon;u) = \frac{\varepsilon_a(u)-\varepsilon}{\Gamma(u)}\cdot\frac{\partial\Gamma(u)}{\partial u_i}-\frac{\partial\varepsilon_a(u)}{\partial u_i}
\end{align}
being the (dynamical) frictional force on the mode $u_i$. This result was derived in Ref. \cite{brandbyge} for a single adsorbate mode and has been generalized to more than one modes here. It is also straightforward to extend the derivation to include $N$ resonant states and the resulting friction is simply the sum of the $N$ partial frictions resulting from each resonance. The diagonal elements of the friction tensor are strictly positive and the main contribution from $\eta_{ij}$ in \eqref{Langevin} will be a frictional force in a direction opposite the velocity. In the presence of hot metallic electrons, the ground state potential appearing in \eqref{Langevin} should actually be replaced by a temperature dependent renormalized potential $V_0(u_i)\rightarrow V_0(u_i)+F(T;u_i)$.\cite{brandbyge} However, the correction is usually so small that it can be neglected and we have explicitly verified this for the systems considered in the present work.

In the present work we will make the Markov approximation where there is no temporal correlation of the fluctuating forces. The approximation is valid when the thermal correlation time $t_c\sim\hbar/k_BT$ is much smaller than the timescale of adsorbate motion, and the fluctuating force $\xi_i(t)$ is a Gaussian distributed stochastic variable with a correlation function given by
\begin{align}\label{correlator}
\langle\xi_i(t_1)\xi_j(t_2)\rangle=2\eta_{ij}k_BT\delta(t_1-t_2).
\end{align}

To summarize, the Langevin equation \eqref{Langevin} can be thought of as describing classical dynamics with stochastic fluctuations. Quantum effects enters through the initial state of the adsorbate and can be included by running classical trajectories with initial conditions sampled from a Wigner distribution of the initial state. For non-quadratic potentials the Langevin equation should be regarded as a semiclassical approximation to the true dynamics. The derivation leading to Eqs. \eqref{Langevin}-\eqref{fric_force} is based on a path integral representation of the reduced density matrix.\cite{brandbyge} In appendix \ref{path} we derive Eq. \eqref{density_harmonic} and show how the Wigner distribution emerges in this formalism and using the technique of Brandbyge et al. \onlinecite{brandbyge} it is straightforward to generalize the result to the full Hamiltonian \eqref{H}.

\subsection{Master equation}
If one is interested in the time dependent probability for the adsorbate to be be in a particular energy eigenstate rather than at certain position, it is more convenient to consider the reduced density matrix in a basis of Hamiltonian eigenstates. Taking the electronic trace of the Liouville equation leads to
\begin{align}\label{liouville}
\frac{d\rho_{red}}{dt}+\frac{i}{\hbar}[H_0,\rho_{red}]=\frac{-i}{\hbar}\mathrm{Tr}_{el}[H_I,\rho],
\end{align}
where $\rho_{red}=\mathrm{Tr}_{el}(\rho)$ is the reduced density matrix and $\mathrm{Tr}_{el}$ is the trace over electronic states. In a basis of eigenstates of $H_0$, the diagonal elements of the reduced density matrix are the time dependent probabilities of finding the adsorbate in a particular state. The right hand side is a complicated functional which depends on the complete history of the density matrix. However, making the self consistent Born approximation, the Markov approximation and neglecting the off-diagonal elements of $\rho_{red}$ leads to the master equation\cite{gao97}
\begin{align}\label{master}
\frac{dp_n}{dt}=\sum_{m=0}^{\infty}\Big(p_mW_{m\rightarrow n}-p_nW_{n\rightarrow m}\Big),
\end{align}
where $p_n=(\rho_{red})_{nn}$ and $W_{m\rightarrow n}$ are the transition rates given by
\begin{align}\label{golden_rule}
W_{m\rightarrow n}=\frac{2\pi}{\hbar}\sum_{q,q'}&n_F(\varepsilon_q)\big(1-n_F(\varepsilon_{q'})\big)|\langle q;m|H_I|q';n\rangle|^2\notag\\
&\times\delta(\varepsilon_q-\varepsilon_{q'}+\varepsilon_n-\varepsilon_m),
\end{align}
where $|q\rangle$ is the eigenstates of $H_{el}$ with eigenenergies $\varepsilon_q$ and $n_F(\varepsilon)$ is the Fermi-Dirac distribution.

\section{Model potentials}\label{model}
As shown above, zero point motion (or any other initial quantum state) can be included in the molecular dynamics by sampling all phase space and weighing each point according to the Wigner distribution of the initial state. For Langevin dynamics this can be tedious work since one has to run a large number of trajectories for each initial point in phase space to get reasonable statistics. An often used approximation to avoid phase space sampling is to use the classical initial conditions which reproduces the energy of the initial quantum state $E_n$. When the friction is small compared to the period of oscillation, one can then use a single initial phase space point with $E_{clas}(x_0,p_0)=E_n$. We will refer to this as the quasiclassical approximation. However, as will be shown below, this method can give rise to seriously misleading results for Langevin dynamics when the timescale of the hot electron pulse is sufficiently short.

\subsection{Quadratic potential}
For a quadratic potential the Langevin equation is exact within second order perturbation theory provided we include the initial quantum state properly. We can thus compare results obtained by integrating the Langevin equation with those obtained from a master equation approach \eqref{master} and transition rates calculated from the Fermi golden rule expression \eqref{golden_rule}. In principle, the two approaches should be equivalent since the level of approximation is the same (Markov approximation and second order perturbation theory) and we can investigate the importance of using quasiclassical initial conditions compared to true quantum initial conditions.

It may be surprising that the classical Langevin equation should give the same result as the master equation which is inherently quantum mechanical. Furthermore, it may not be obvious how the probabilities $p_n$, which is the basic quantity calculated within the master equation approach, can be extracted from Langevin dynamics. However, if one has access to the Wigner distribution $\mathcal{P}(x,p)$ at a given time, it is indeed possible to calculate $p_n$ since
\begin{align}
p_n&=\langle n|\rho|n\rangle=\int dxdy\rho(x,y)\varphi^*_n(x)\varphi_n(y)\\
&=\int dudv\rho(u+v/2,u-v/2)\notag\\
&\quad\times\int d\tilde{v}\varphi^*_n(u+\tilde{v}/2)\varphi_n(u-\tilde{v}/2)\delta(v-\tilde{v})\notag\\
&=\int dudv\rho(u+v/2,u-v/2)\notag\\
&\quad\times\frac{1}{2\pi\hbar}\int d\tilde{v}dp\varphi^*_n(u+\tilde{v}/2)\varphi_n(u-\tilde{v}/2)e^{ip(v-\tilde{v})/\hbar}\notag\\
&=2\pi\hbar\int dudp\mathcal{P}_n(u,p)\mathcal{P}(u,p),\notag
\end{align}
where $\mathcal{P}_n(u,p)$ is the Wigner distribution of the pure state density matrix $\rho_n=|n\rangle\langle n|$. Integrating the Langevin equation gives rise to a final state classical phase space distribution, but since the equation of motion for a classical phase space distribution is identical to that of a Wigner distribution in a harmonic potential,\cite{hillery} we can identify the final state classical phase space distribution with the final state Wigner distribution.

\begin{figure}[tb]
          \includegraphics[width=8.5 cm, clip]{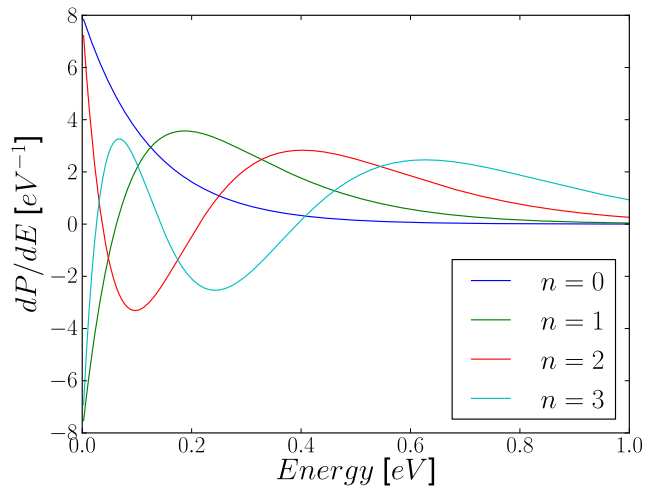}
\caption{The energy distributions given by Eq. \eqref{P_clas^n} for the lowest four vibrational states of a harmonic oscillator with zero point energy $E_0=0.125\;eV$. The corresponding quasiclassical distributions are deltafunctions centered at $E_0(2n+1)$.}
\label{fig:laguerre}
\end{figure}

The pure state Wigner distributions in a quadratic potential is given by\cite{hillery}
\begin{align}
\mathcal{P}_n(x,p)=\frac{(-1)^n}{\pi\hbar}e^{-\mathcal{H}(x,p)/E_0}L_n(2\mathcal{H}(x,p)/E_0),
\end{align}
where $\mathcal{H}(x,p)=p^2/2m+m\omega^2x^2/2$ is the classical Hamiltonian, $E_0=\hbar\omega/2$, and $L_n$ is the $n$'th Laguerre polynomial. Since $\mathcal{P}_n$ is only a function of the Hamiltonian energy we can write
\begin{align}\label{p_n}
p_n&=2\pi\hbar\int_0^\infty dE\mathcal{P}_n(E)\frac{dP}{dE}\\
&=2(-1)^n\int_0^\infty dEe^{-E/E_0}L_n(2E/E_0)\frac{dP}{dE},\notag
\end{align}
with
\begin{align}\label{P_clas}
\frac{dP}{dE}=\int dxdp\mathcal{P}(x,p)\delta(E-\mathcal{H}(x,p)).
\end{align}
Note that the distribution $dP/dE$ is not a true probability distribution since it is not strictly positive, but it can be rigorously translated into the quantum mechanical probabilities $p_n$. On the other hand, we can obtain the distribution $dP_n/dE$ associated with a particular vibrational state $|n\rangle$ by replacing $\mathcal{P}(x,p)$ in Eq. \eqref{P_clas} with $\mathcal{P}_n(x,p)$. Using that $dxdp=\hbar d\varphi d\mathcal{H}/2E_0$ with $\varphi$ being a phase space angle, the integral can then be evaluated to
\begin{align}\label{P_clas^n}
\frac{dP_n}{dE}=\frac{(-1)^n}{E_0}e^{-E/E_0}L_n(2E/E_0).
\end{align}
The distributions Eq. \eqref{P_clas^n} are shown in Fig. \ref{fig:laguerre} for the first four vibrational states with $E_0=0.125\;eV$. The structure of the distributions is in sharp contrast to that obtained in the quasiclassical (QC) approach where the energy is fixed at $E_n$ and the energy distribution of the $n$'th state is $dP_n^{(QC)}/dE=\delta(E-E_n)$ with
$E_n=\hbar\omega(n+1/2)$. This gives rise to completely different and and even negative probabilities. For example, using $dP_0^{(QC)}/dE=\delta(E-E_0)$ immediately yields $p_0=p_1=-p_2=0.74$ from Eq. \eqref{p_n}.

\begin{figure}[tb]
          \includegraphics[width=4.2 cm, clip]{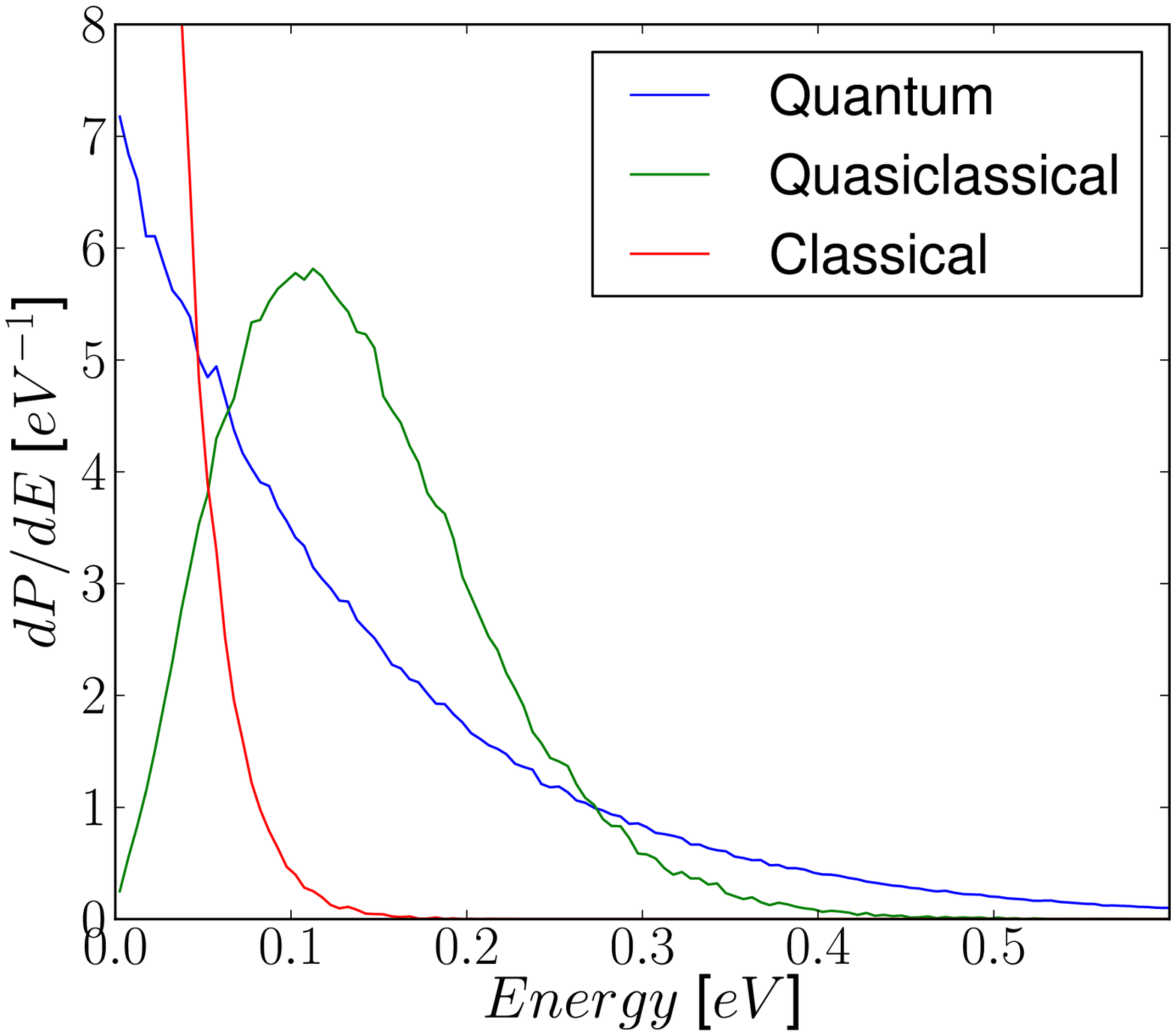}
          \includegraphics[width=4.2 cm, clip]{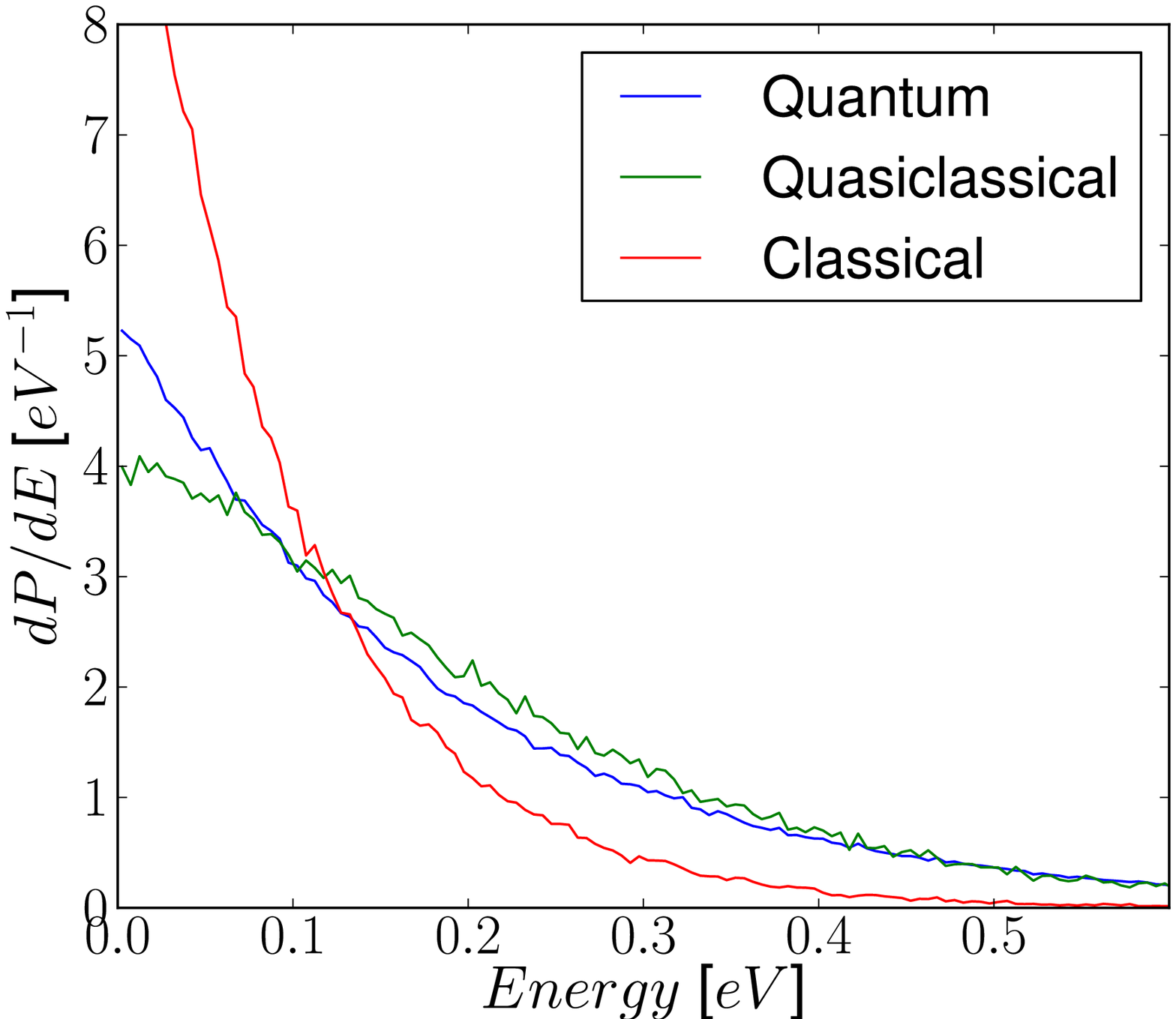}
\caption{The continuous energy distributions $dP/dE$ obtained from Langevin dynamics with a constant $T_e=4000\;$ using quantum, quasiclassical, and classical boundary conditions. The initial quantum state is the vibrational ground state. Left: $t=0.1\;ps$. Right: $t=0.5\;ps$. After a while both the quasiclassical and classical distributions approach the quantum distribution.}
\label{fig:dist_time}
\end{figure}

We have performed Langevin dynamics using Eqs. \eqref{Langevin} and \eqref{friction} with a single mode and a linear interaction Hamiltonian: $H_I=-fc_a^\dag c_ax$ using the parameters $m=	6.86\;amu$, $\hbar\omega=0.25\;eV/$, $\varepsilon_0=2.6\;eV$, $\Gamma=2.0\;eV$, and $f=8.7\;eV/$\textit{\AA}. These parameters were chosen to mimic the internal vibrational mode of CO adsorbed on Cu(100) considered below, but presently we will just think of them as a realistic set of parameters which we use to compare different model calculations. The adsorbate is initially in its ground state described by the Wigner distribution
\begin{align}
\mathcal{P}_0(x_0,p_0)=\frac{1}{\pi\hbar}e^{-x_0^2/x_Q^2-p_0^2/p_Q^2}
\end{align}
with the quantum length and momentum given by
\begin{align}
x_Q=\sqrt{\hbar/m\omega}, \qquad p_Q=\sqrt{\hbar m\omega}.
\end{align}
The distribution is even in both momentum and position and since the frictional decay is much slower than the vibrational time of oscillation, the final state phase space distribution can be assumed to be even in the initial phase space point. For simplicity we assume a constant electronic temperature at $T_e=4000\;K$ and integrate the Langevin equation for $t=1\;ps$. For each point on an initial (6x6) positive phase space grid with a spacing $0.5x_Q\times0.5p_Q$, we run a large number of Langevin trajectories ($\sim30000$) and record the final state energy. The final state energy distribution is then obtained by summing the distributions resulting from each initial phase space point $dP/dE(E;x_0,p_0)$ weighted by the initial state Wigner distribution $\mathcal{P}(x_0,p_0)$: 
\begin{align}\label{dPdE}
\frac{dP(E)}{dE}=\int dx_0dp_0\mathcal{P}(x_0,p_0)\frac{dP(E;x_0,p_0)}{dE}.
\end{align}
In Fig. \ref{fig:dist_time} we show this distribution at $t=0.1\;ps$ and $t=0.5\;ps$ along with the distributions resulting from quasiclassical (initial phase space points with $\mathcal{H}(x_0,p_0)=E_0$) and classical initial condition (initial phase space point $x_0=p_0=0$).
On long time scales the distributions will forget the initial conditions and approach a Boltzmann distribution at the appropriate temperature. However, on timescales less than a picosecond there is still plenty of memory of the initial state and the classical and quasiclassical distributions, which start as delta functions at $E=0$ and $E=E_0$ respectively, are completely wrong at timescales on the order of $0.1\;ps$. The quasiclassical initial conditions approach the correct distribution faster than the classical one since the initial state contains the right amount of energy which just needs to be redistributed.

\begin{figure}[tb]
          \includegraphics[width=4.2 cm, clip]{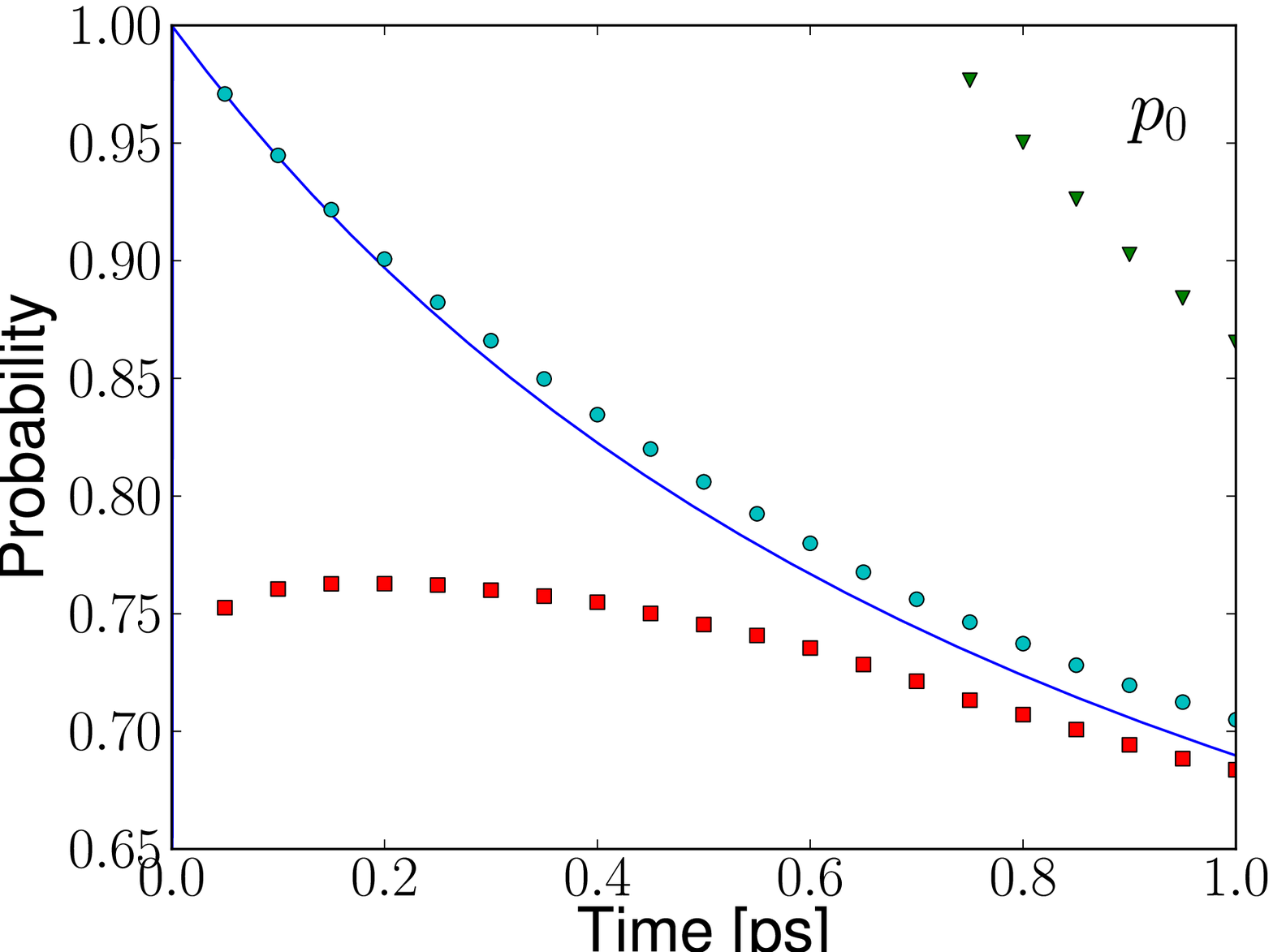}
          \includegraphics[width=4.2 cm, clip]{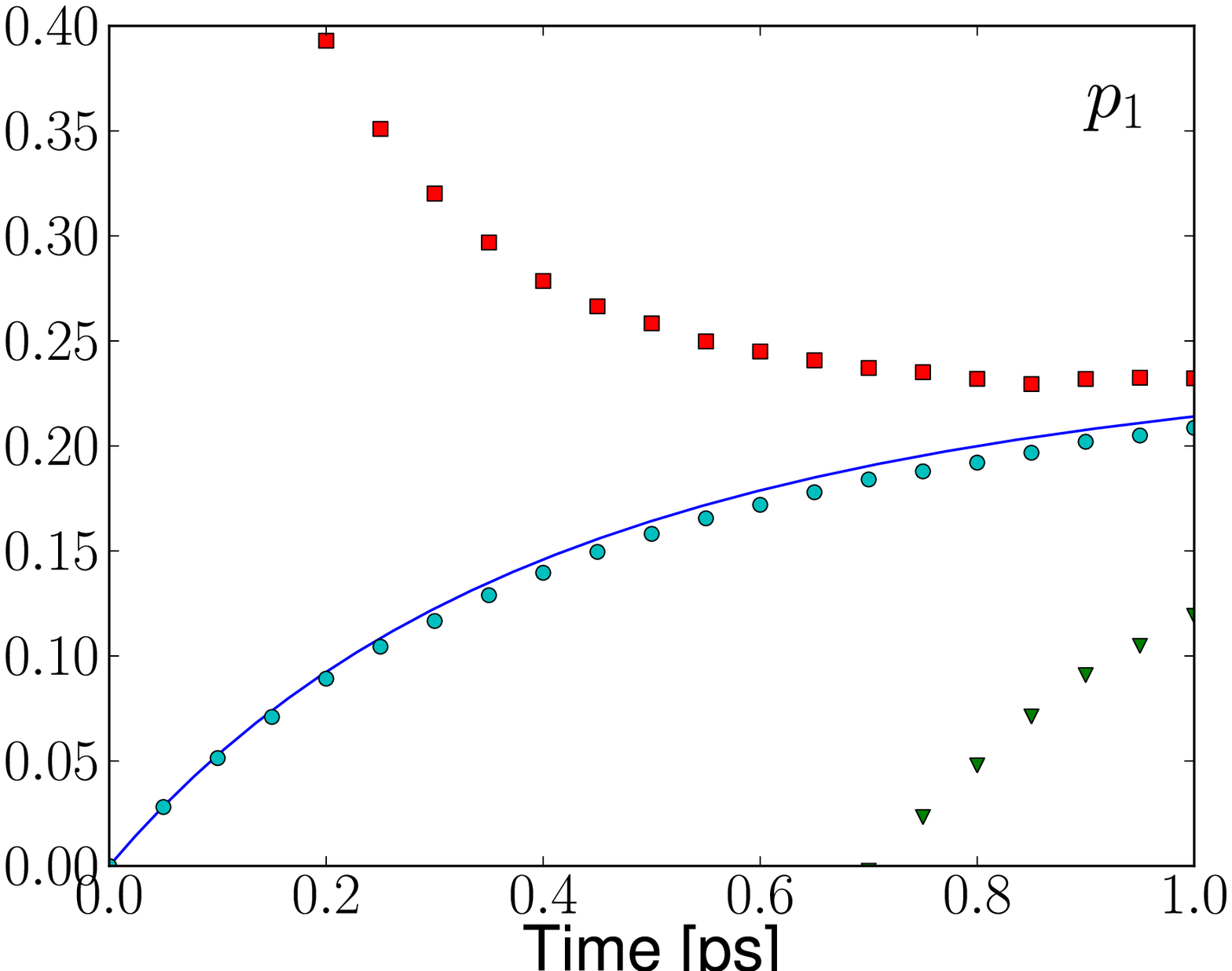}
          \includegraphics[width=4.2 cm, clip]{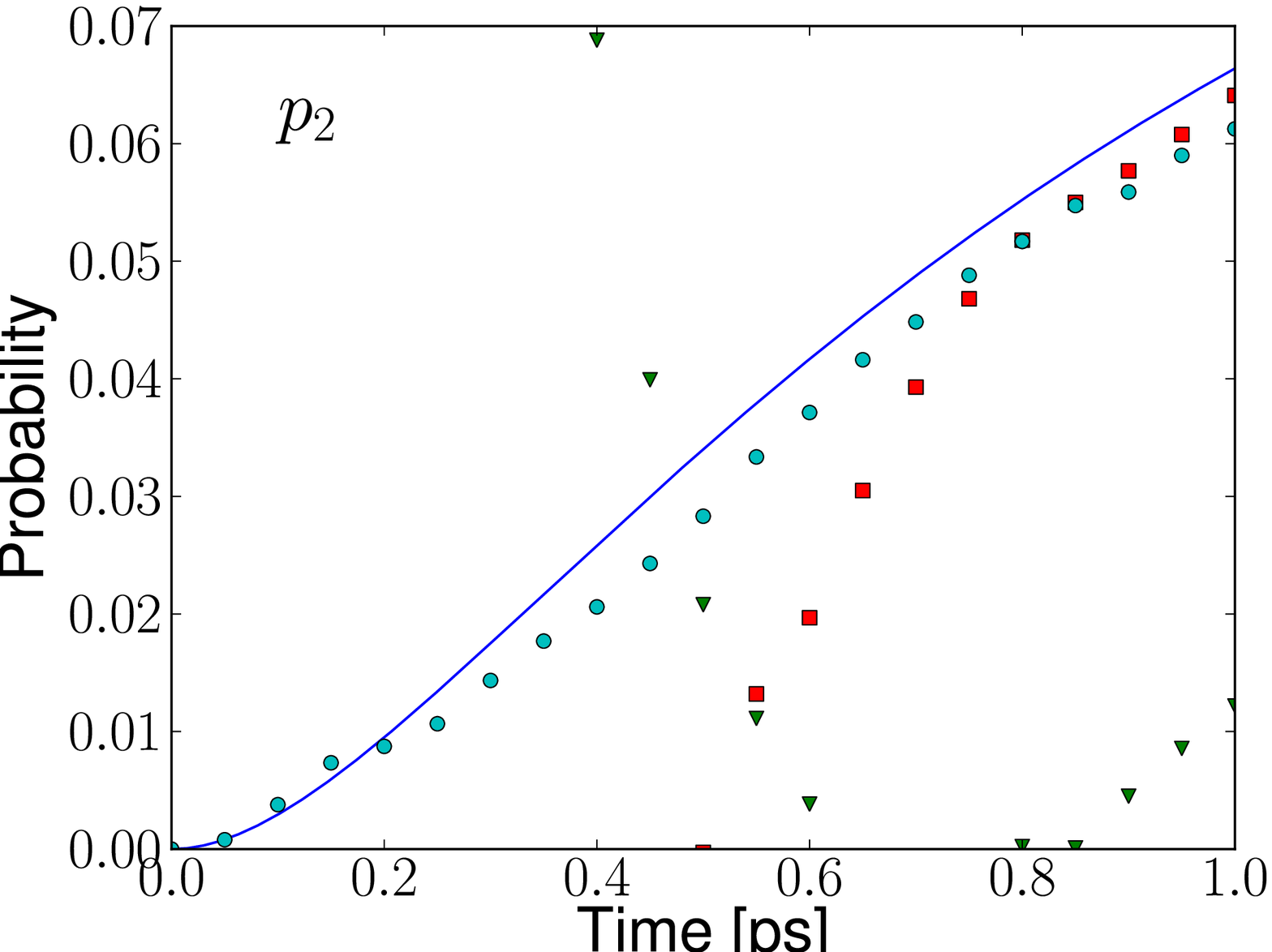}
          \includegraphics[width=4.2 cm, clip]{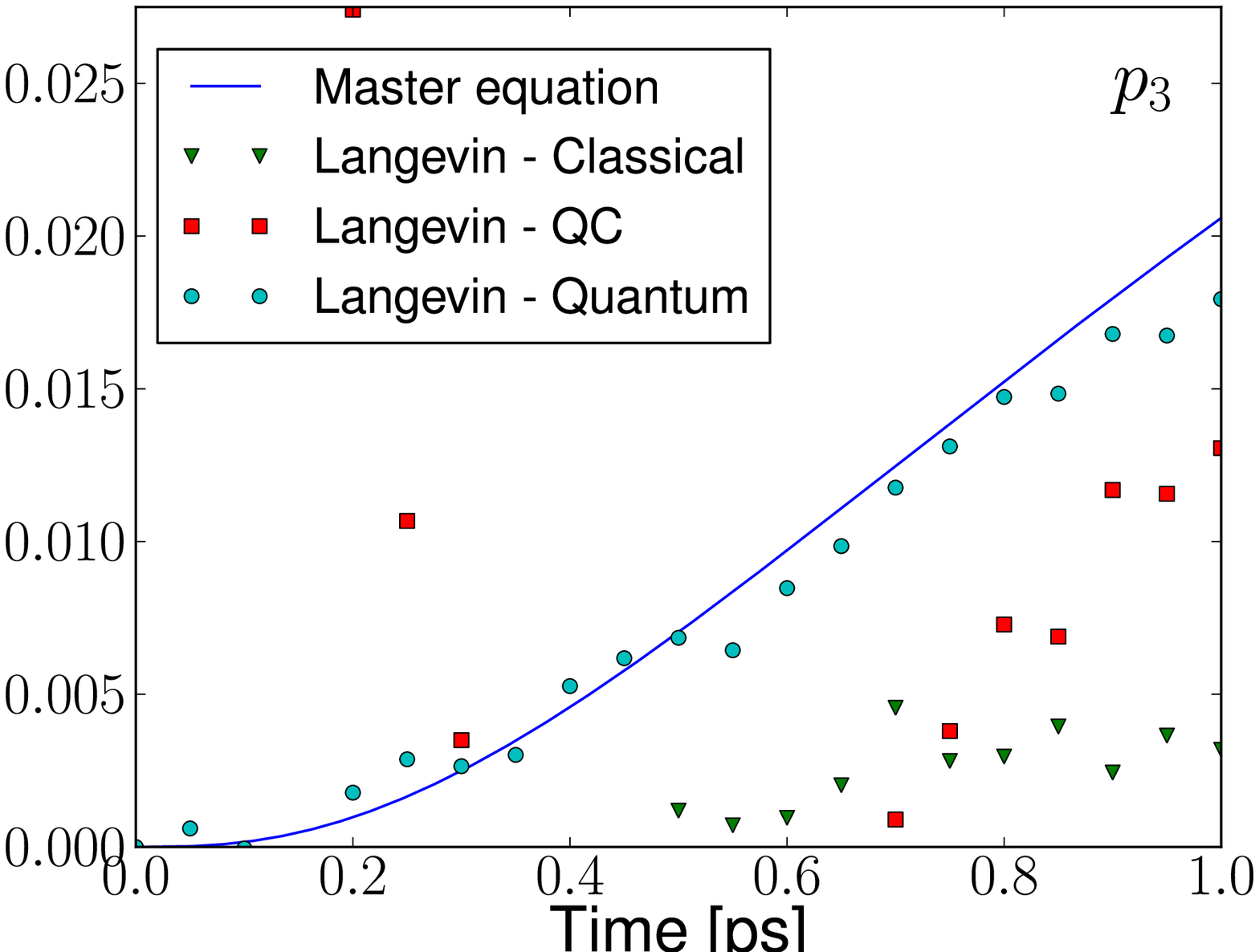}
\caption{The time dependent probabilities $p_n$ for being in the vibrational state $|n\rangle$ obtained with the master equation and Langevin dynamics with three kinds of initial conditions. The correct quantum initial conditions are seen to give results nearly identical to the master equation, whereas the classical and quasiclassical initial conditions give wrong results. For small time scales the classical and quasiclassical initial conditions are not shown since the are not consistent with the harmonic oscillator Wigner distribution in the sense that they give rise to probabilities which are negative or larger than one.}
\label{fig:p_n}
\end{figure}

With the interaction Hamiltonian $H_I=-fc_a^\dag c_ax$ it is easy to calculate the transition rates Eq. \eqref{golden_rule} with the result:
\begin{align}
W_{m\rightarrow n}=&m\delta_{m,n+1}\frac{\pi f^2}{M\omega}\int d\varepsilon\rho_a(\varepsilon)\rho_a(\varepsilon+\hbar\omega)\notag\\
&\times n_F(\varepsilon)\big(1-n_F(\varepsilon+\hbar\omega)\big)\notag\\
+&(m+1)\delta_{m,n-1}\frac{\pi f^2}{M\omega}\int d\varepsilon\rho_a(\varepsilon)\rho_a(\varepsilon-\hbar\omega)\notag\\
&\times n_F(\varepsilon)\big(1-n_F(\varepsilon-\hbar\omega)\big).
\end{align}
Using the parameters above we can then integrate the master equation Eq. \eqref{master} and compare the probabilities $p_n$ with those obtained from the Langevin equation Eqs. \eqref{p_n} and \eqref{dPdE}. This is shown in Fig. \ref{fig:p_n} for the four lowest vibrational states. As expected we see a close correspondence between the master equation approach and Langevin dynamics with correct phase space sampling. In contrast, the classical initial conditions result in completely wrong probabilities and the quasiclassical initial conditions only result in sensible probabilities after $\sim0.5\;ps$. 

It should be noted, that the quasiclassical initial conditions gives a good description of average quantities and the average energy $\langle E\rangle=\sum_np_nE_n$ is very well approximated by the quasiclassical approach, even at short timescales. However, if one were to model a surface reaction with a barrier by a truncated harmonic potential the quasiclassical approach is likely to fail. For example, the adsorption energy of CO on Cu(100) is $\sim0.6\;eV$ and as a simple model for hot electron induced desorption one could use the present oscillator truncated above the desorption energy. This means that $p_2+p_3$ would be a measure of the desorption probability and from Fig. \ref{fig:p_n} it is clear that for times $<0.5\;ps$ one would severely miscalculate the desorption probability.

\subsection{Morse potential}
Allthough the quadratic potential comprises a nice toy model for comparing Langevin dynamics with the master equation approach, it is not particularly well suited to simulate surface reactions such as desorption or dissociation. We will make a simple model for a desorption potential and modify the quadratic potential considered above to a one-dimensional Morse potential $V_M(x)=D(1-e^{-ax})^2$ with $D=0.57\;eV$. The parameter $a$ is determined by requiring that the second derivative at the minimum of the well match the frequency of the harmonic potential considered above. A quantization of this potential yields five bound states with energies $E_n$ and a continuum of free states with energies $E_k=\hbar^2k^2/2m$. 

Under the influence of a thermal pulse of electrons, a bound state $|m\rangle$ can make transitions to other bound states $|n\rangle$ or to free states $|k\rangle$. The transition rates can be calculated within second order perturbation theory and the result is
\begin{align}
W_{m\rightarrow n}=&\frac{2\pi f^2|\langle m|x|n\rangle|^2}{\hbar}\int d\varepsilon\rho_a(\varepsilon)\rho_a(\varepsilon+\hbar\omega_{mn})\notag\\
&\times n_F(\varepsilon)\big(1-n_F(\varepsilon+\hbar\omega_{mn})\big)
\end{align}
for bound state transitions and 
\begin{align}
W_{m\rightarrow k}=&\frac{2\pi f^2|\langle m|x|k\rangle|^2}{\hbar}\int d\varepsilon\rho_a(\varepsilon)\rho_a(\varepsilon+\hbar\omega_{mk})\notag\\
&\times n_F(\varepsilon)\big(1-n_F(\varepsilon+\hbar\omega_{mk})\big)
\end{align}
for transitions to free states. Here we have defined $\hbar\omega_{mi}=E_m-E_i$. The matrix elements have been calculated previously\cite{lima} and it is now straightforward to integrate the master equation \eqref{master}. We will interpret the probability of being in a free state $|k\rangle$ at time $t$ as the desorption probability.

For a non-quadratic potential the Langevin equation is based on a semiclassical approximation. However, since the master equation \eqref{master} is still correct within second order perturbation theory we can explicitly examine the validity of the semiclassical approximation by comparing the two approaches. Due to the lack of a classical/quantum correspondence for the Morse potential it is not possible to convert the classical energy distribution resulting from Langevin dynamics into probabilities of being in eigenstates of the Morse potential. Nevertheless, it is natural to associate the probability of being in a continuum state $|k\rangle$ with the probability that a classical trajectory results in a final state energy $E_k$. The initial quantum state is included as described above by sampling phase space and integrate weighting by the Wigner distribution. The Wigner distribution of the Morse potential ground state is well known\cite{bund}, but since it is not even in the position coordinate we need to sample twice the phase space compared with the harmonic oscillator.

\begin{figure}[tb]
          \includegraphics[width=4.2 cm, clip]{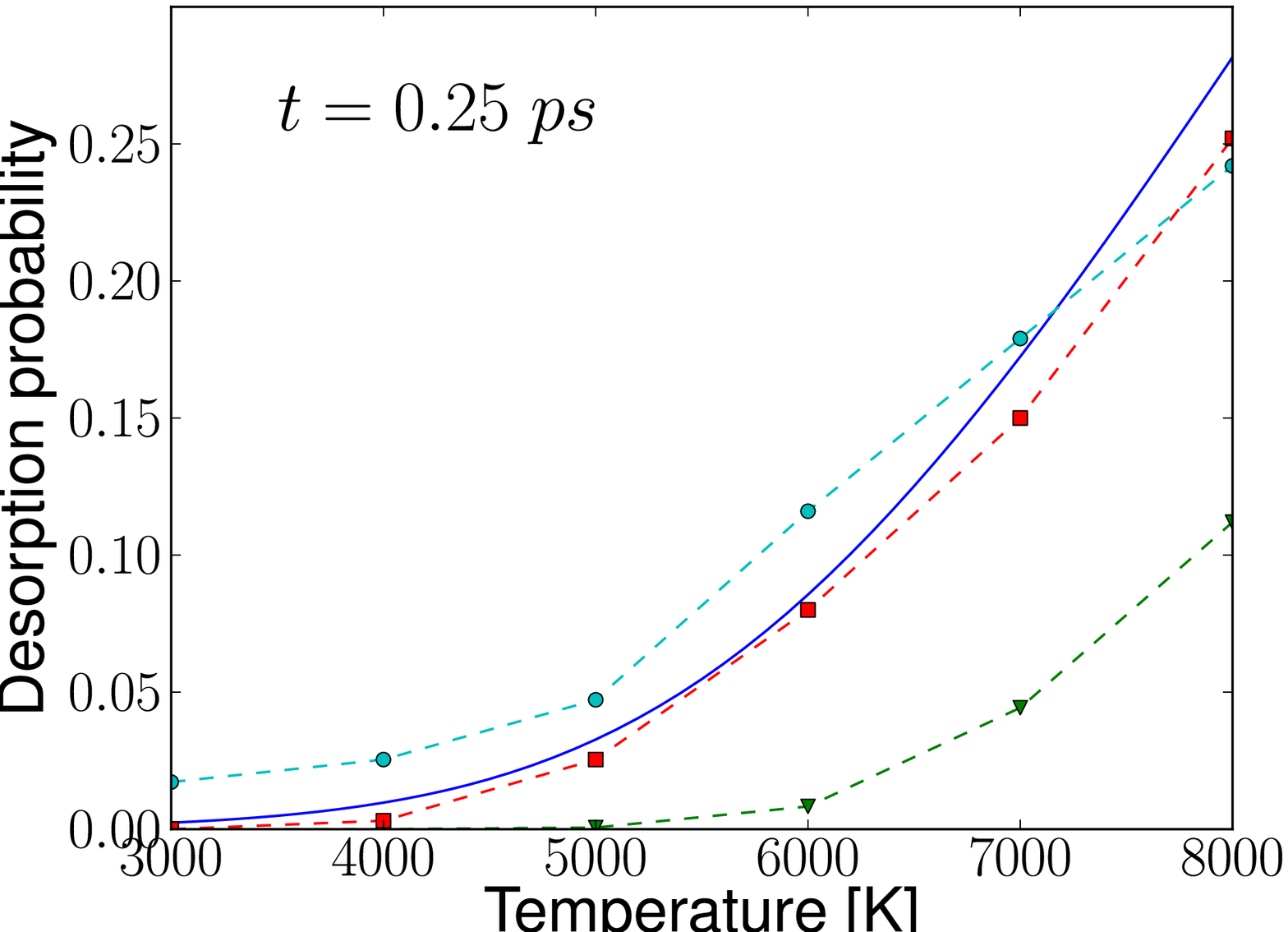}
          \includegraphics[width=4.2 cm, clip]{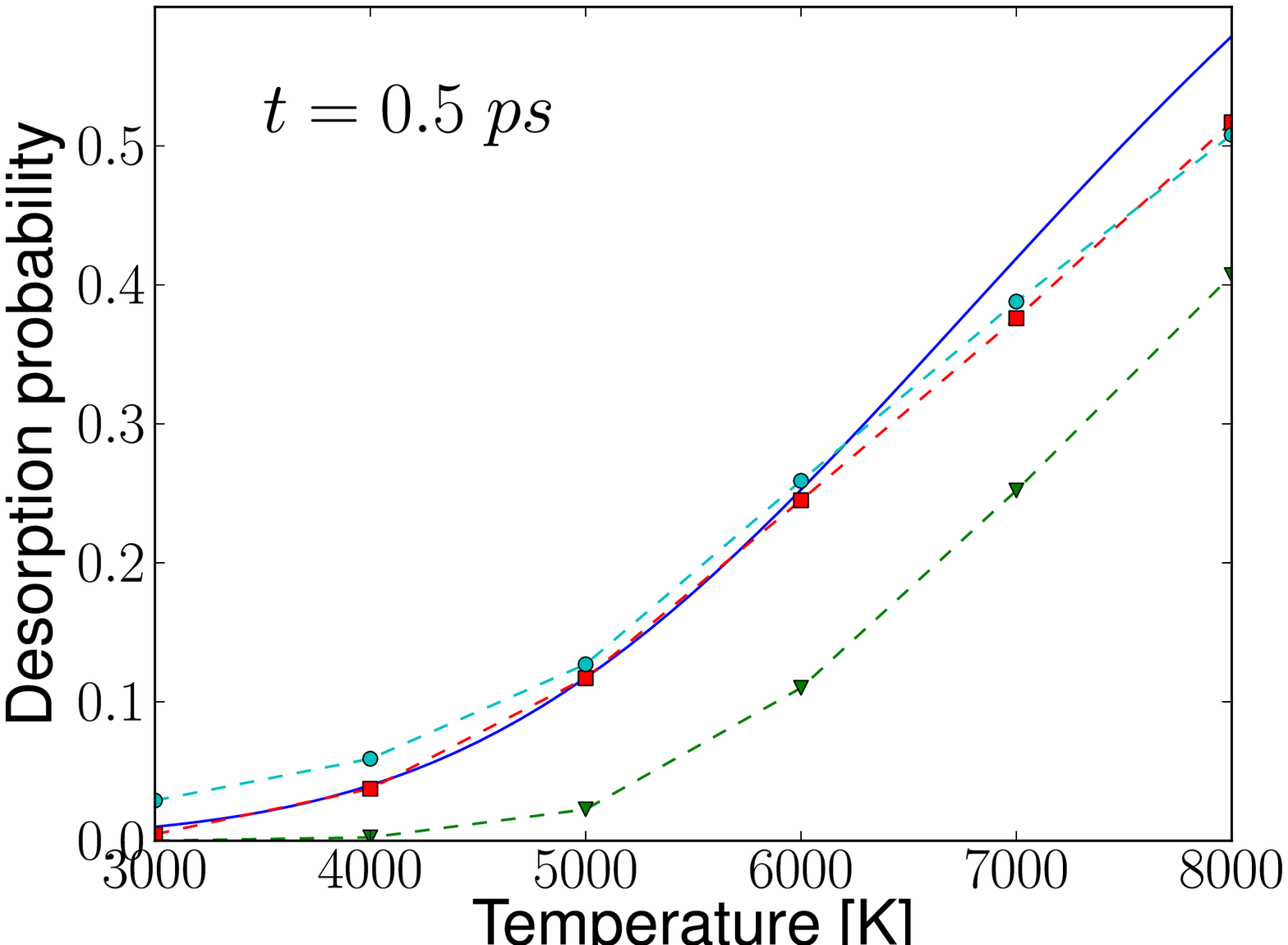}
          \includegraphics[width=4.2 cm, clip]{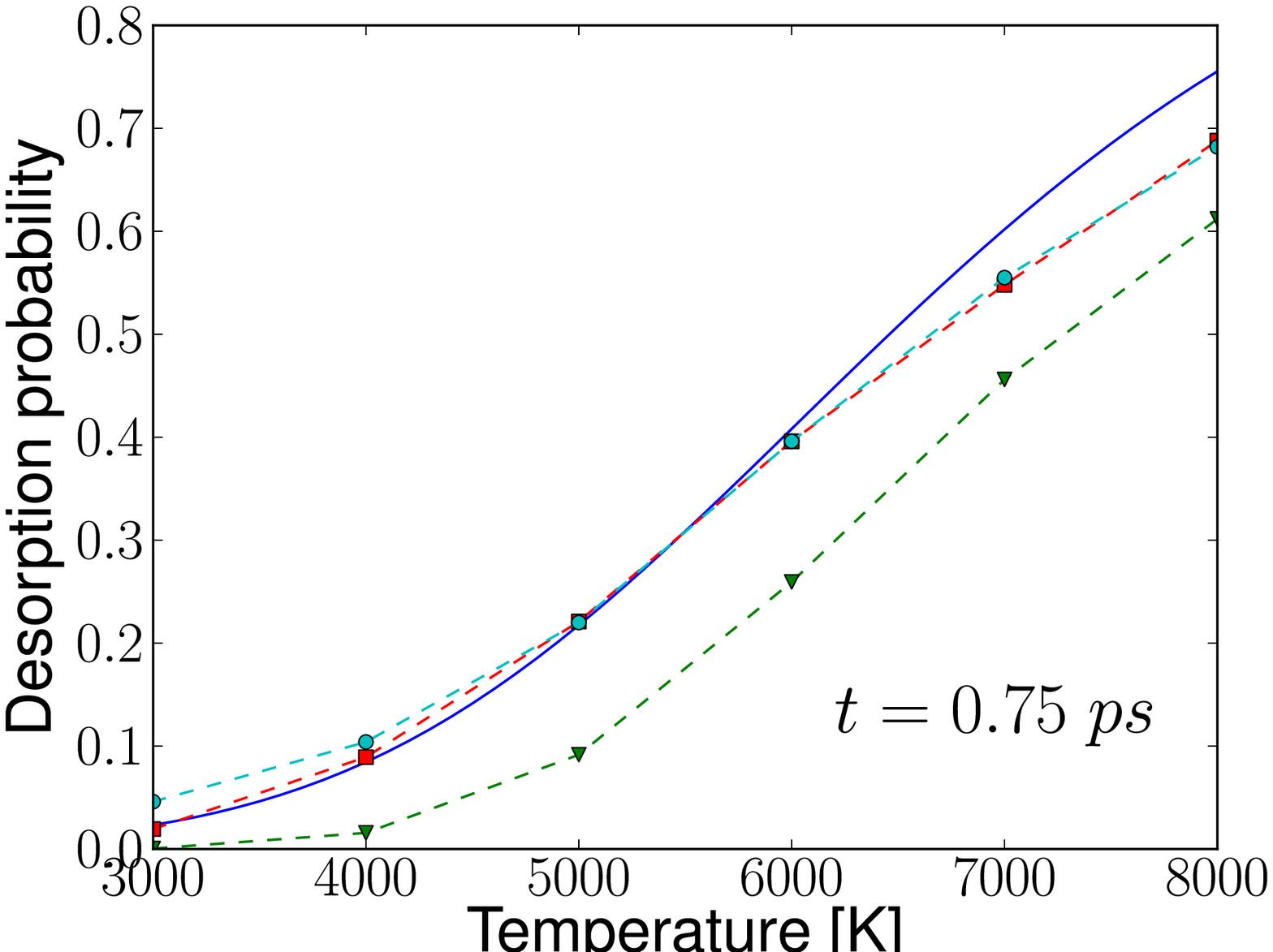}
          \includegraphics[width=4.2 cm, clip]{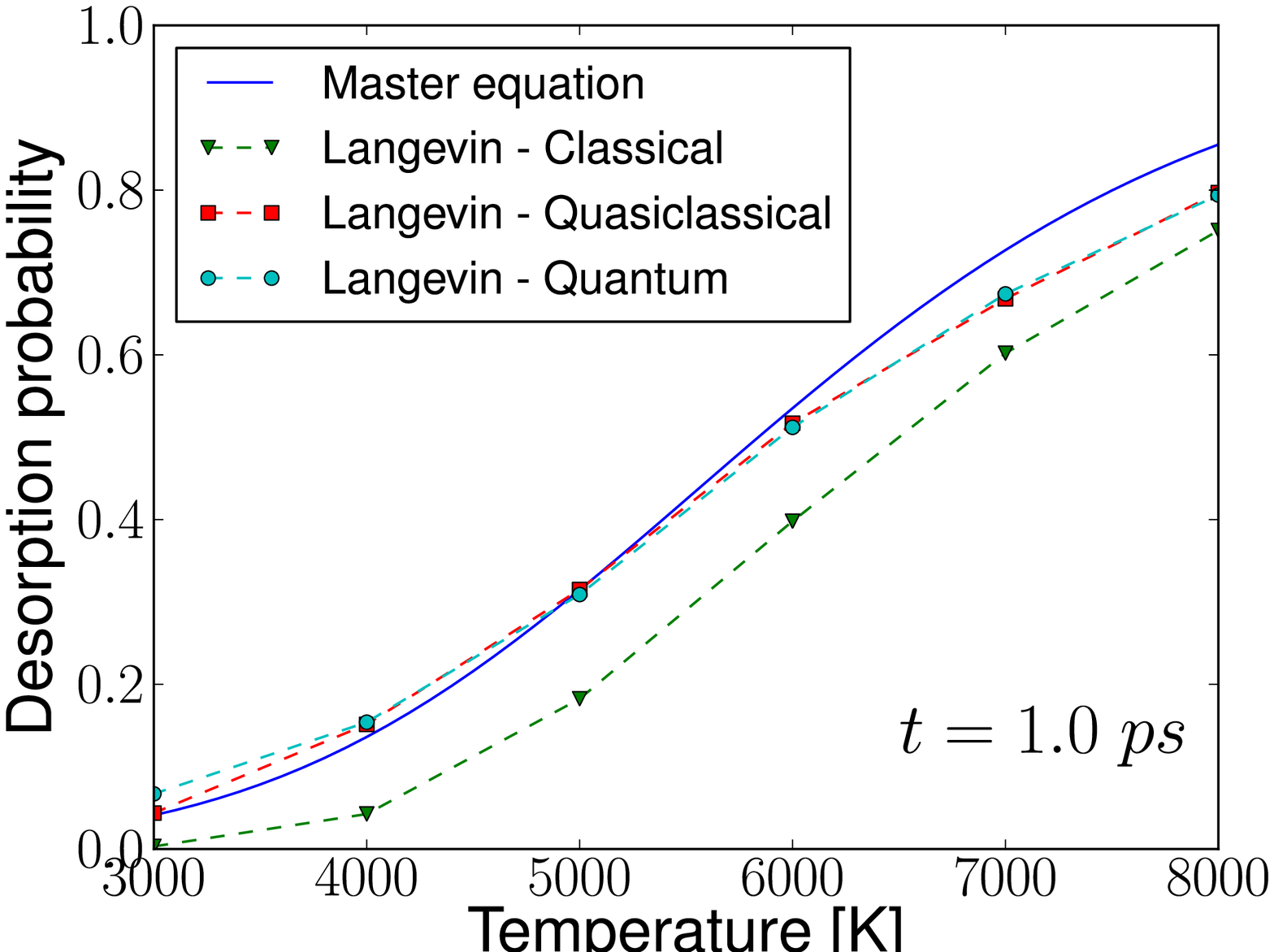}
\caption{Desorption probabilities as a function of the electronic temperature $T_e$ calculated from the master equation approach and Langevin dynamics with classical quasiclassical and quantum initial conditions. The four figures show the desorption probability after interaction times of $0.25\;ps$, $0.5\;ps$, $0.75\;ps$, and $1.0\;ps$ respectively.}
\label{fig:morse}
\end{figure}
The desorption probabilities calculated with the master equation and Langevin dynamics is shown in Fig. \ref{fig:morse}. For $t=0.25\;ps$, the probabilities show significant deviation signalling a breakdown of classical time evolution at small time scales which is expected. It is a bit more surprising, that the high temperature limit deviates from the quantum probabilities even at $t=1\;ps$. This could be due a breakdown of perturbation theory at such high temperatures, since the effective perturbation of the system becomes large when the electronic temperature is increased. We also show the probabilities resulting from Langevin dynamics with classical and quasiclassical initial conditions and it is again seen that the classical initial conditions severely underestimates the probabilities. In contrast to the harmonic oscillator, the quasiclassical approach is in very good approximation for the quantum initial conditions when calculating desorption probabilities. This is due to the fact that the quasiclassical approach is a good approximation for average quantities and the desorption probability in the present case is an integral over a continuum of excited states $|k\rangle$. This will be extremely useful since the quasiclassical approximation allows us to circumvent phase space sampling.

\section{Ab initio potential}\label{ab_initio}
\begin{figure}[tb]
          \includegraphics[width=8.5 cm, clip]{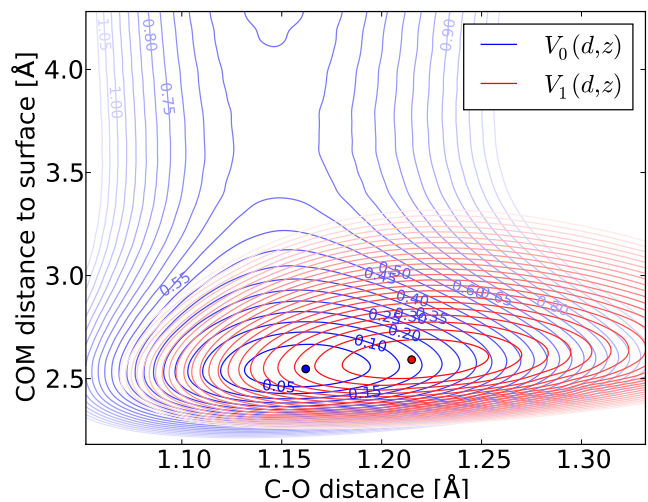}
\caption{Potential energy surfaces for the ground and excited state of CO adsorbed at a Cu(100) top site. The contours are at 0.05 eV intervals and the desorption barrier is at 0.57 eV. The extra electron in the anti-bonding $2\pi$ orbital is seen to stretch th C-O bond. The center of mass is moved slightly out from the surface in spite of the attraction to the image charge.}
\label{fig:pes}
\end{figure}
As an example illustrating quantum effects in Langevin dynamics using ab initio potentials, we consider CO adsorbed on Cu(100). This system has previously been investigated in the context of electronic friction and the closely connected vibrational linewidth broadening induced by electron hole pair excitations.\cite{persson, head-gordon, tully} All parameters in the model Hamiltonian \eqref{H} was obtained within Density Functional Theory (DFT) using the code \texttt{gpaw},\cite{gpaw, mortensen} which is a real-space Density Functional Theory (DFT) code that uses the projector augmented wave method.\cite{blochl1,blochl2} We used a grid spacing of $0.2$ \textit{\AA} and the calculations were performed in a (2x2) supercell sampled by a (4x6) grid of $k$-points using the RPBE\cite{hammer} exchange correlation functional. The system was modelled by a three layer Cu(100) slab with the top layer relaxed and CO adsorbed in a c(2x2) structure (0.5 coverage at top sites). For this system the electronic friction is dominated by the unoccupied $2\pi$ orbitals which we assume to represent the resonant state $|a\rangle$. 

We have calculated the potential energy surfaces in terms of the center of mass and bond length coordinates which are denoted by $z$ and $d$ respectively. We restrict the analysis to these modes since in a first order Taylor expansion of $\varepsilon_a(x_i)$, the frustrated rotations and translations do not couple to the resonant state due to symmetry. The desorption energy is determined to be $E_{des}\sim 0.57\;eV$ in excellent agreement with the experimental value \cite{struck}. The excited state potential energy surface $V_1(d,z)$ was calculated using a generalization of the $\Delta$-self-consistent field method where the resonant state is expanded in a basis of Kohn-Sham orbitals and the resulting resonant density is added to the density in each iteration step. Thus for each adsorbate position we calculate the energy resulting from forcing an electron into a $2\pi$ orbital which is then not an eigenstate of the full electronic system. The excited state thus has a finite lifetime which in the wide band limit can be related to the resonance width as $\tau=\hbar/\Gamma$.\cite{olsen2} For details on the method and comparison with experiments we refer to.\cite{gavnholt} Since electrostatic interactions may arise between an excited molecule and its periodic image we have checked that the excited state calculations do not change significantly when the supercell is changed to (4x4). 

\begin{figure}[t]
	\includegraphics[width=8.5 cm]{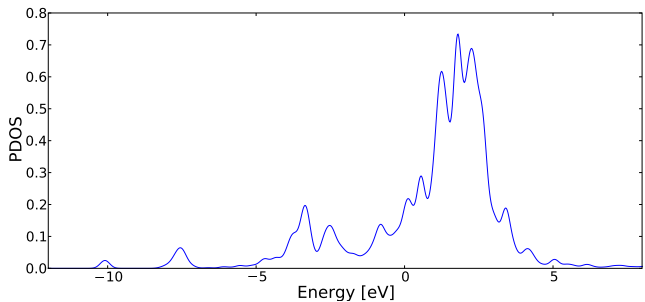}
\caption{Density of states projected onto the $2\pi$ orbital of CO adsorbed on Cu(100) top site. The full width at half maximum is estimated to be $\Gamma=2.0\;eV$. The Fermi level is at $E=0\;eV$ and the resonance is seen to be mostly unoccupied in the electronic ground state.}
\label{fig:pdos}
 \end{figure}
The ground and excited state potential energy surfaces are shown in figure \ref{fig:pes}. The ground state is well approximated by a quadratic potential in the internal mode and a Morse potential in the center of mass mode. The two modes are nearly decoupled and in Tab. \ref{tab1} we display the parameters associated with the two modes at the ground state minimum. The resonance width $\Gamma$ was obtained from the projected density of states shown in figure \ref{fig:pdos}. At the ground state equilibrium position the width is approximately $\Gamma_0\approx2\;eV$ and varying the adsorbate position shows that the coordinate dependence is well approximated by $\Gamma=\Gamma_0e^{-z/z_{\Gamma}}$ with $z_{\Gamma}\approx0.7$ \textit{\AA}. Since the friction tensor is additive in contributing orbitals, we can simply multiply the expression \eqref{friction} by a factor of four to account for the degeneracy of the $2\pi$ orbital and spin, or equivalently, multiply the frictional force by a factor of two which for the internal mode reproduces the parameters used in Sec. \ref{model}. The excitation energy at the ground state minimum is $\varepsilon_0=2.6\;eV$. The diagonal elements of the friction tensor Eq. \eqref{friction} at the equilibrium position and zero temperature can be roughly related to the vibrational lifetimes of the modes: $\tau_i=M_i/\eta_{ii}$.
\begin{figure}[tb]
          \includegraphics[width=8.5 cm, clip]{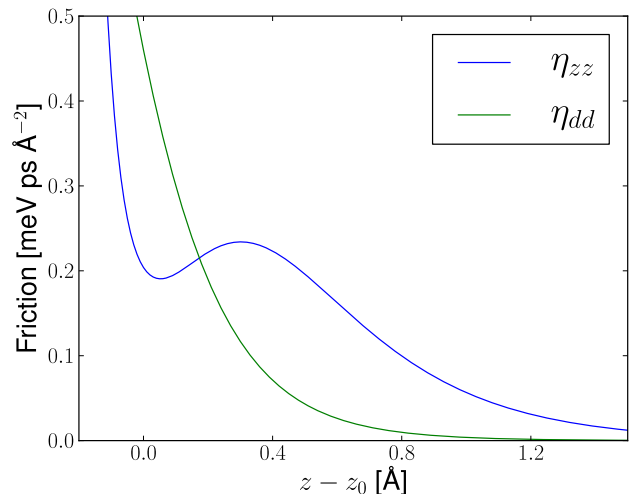}
\caption{Diagonal components of the friction tensor as a function of COM distance to surface evaluated at $T=6000\;K$. Both components decrease exponentially far from the surface but have very different behavior near the minimum position.}
\label{fig:friction}
\end{figure}
In Fig. \ref{fig:friction} we show the two diagonal components as a function of distance to the surface. The two components have the same order of magnitude near the equilibrium position ($z-z_0=0$), but the friction in the internal mode ($\eta_{dd}$) is seen to decay much faster far from surface than the COM friction. Furthermore, the COM friction has a local maximum beyond the equilibrium position and the molecule is thus likely to dissipate energy on the path leading to desorption which decreases the desorption probability. It should be noted that although the frictional force parameters $f_i$ have the same order of magnitude, they originate from different terms in Eq. \eqref{fric_force}. The center of mass minimum is nearly unaffected by a transition to the excited state as seen in figure \ref{fig:pes} and the frictional force arises only from the COM dependence of the resonance width. On the other hand, the resonance width is nearly independent of the internal stretch mode and the internal frictional force originate in the large displacement of the excited state minimum position. The vibrational lifetimes are in good agreement with previous calculations using a different method.\cite{persson, head-gordon, tully}
\begin{table}[t]
\begin{center}
\begin{tabular}{c|c|c|c}
 & $\hbar\omega_i$ & $f_i(\varepsilon_F)$ & $M_i/\eta_{ii}(0;0)$ \\
	\hline
Internal & 0.248 eV & 4.3 eV/\AA & 2.7 ps\\
COM      & 0.043 eV & -3.6 eV/\AA & 16 ps\\
\end{tabular}
\end{center}
\caption{Parameters for the internal vibration and center of mass mode for CO adsorbed on a Cu(100) top site.}
\label{tab1}
\end{table}

To model a particular surface experiment where a femtosecond laser pulse induces a surface reaction, one would need a detailed model for the time dependent distribution of hot electrons resulting from the laser pulse. In the present paper we do not aim at a precise quantitative calculation of reaction rates, but rather wish to examine the qualitative impact of including quantum initial states in the dynamics. Therefore, we will take a very simple model for the hot electrons and assume a thermal pulse with a Gaussian temporal shape $T_e(t)=T_{max}e^{-t^2/2\Delta t^2}$ with $T_{max}=4000\;K$ and $\Delta t=0.5\;ps$. Under the influence of this pulse we have performed Langevin dynamics with classical quasiclassical and quantized initial conditions in both the internal and center of mass mode using the potentials shown in Fig. \ref{fig:pes}. The Langevin equation is integrated from $2\;ps$ prior to the center of the pulse to $4\;ps$ after the center of the pulse. Due to the very weak coupling between the two modes the initial condition of the internal mode has almost no influence on desorption probabilities. With fully quantized initial conditions (vibrational ground state) of the COM mode we find a desorption probability of $P_{Quan}=3.7\times10^{-6}$, whereas we find $P_{QC}<\times10^{-6}$ and $P_{Clas}<10^{-6}$ when using quasiclassical and classical initial conditions respectively ($10^6$ trajectories did not result in a single desorption event). We note, that when calculating the fluctuating forces Eq. \eqref{correlator}, it is most important to take into account the correlation between the two modes determined by the off-diagonal elements of the friction tensor.

\begin{figure}[tb]
          \includegraphics[width=4.25 cm, clip]{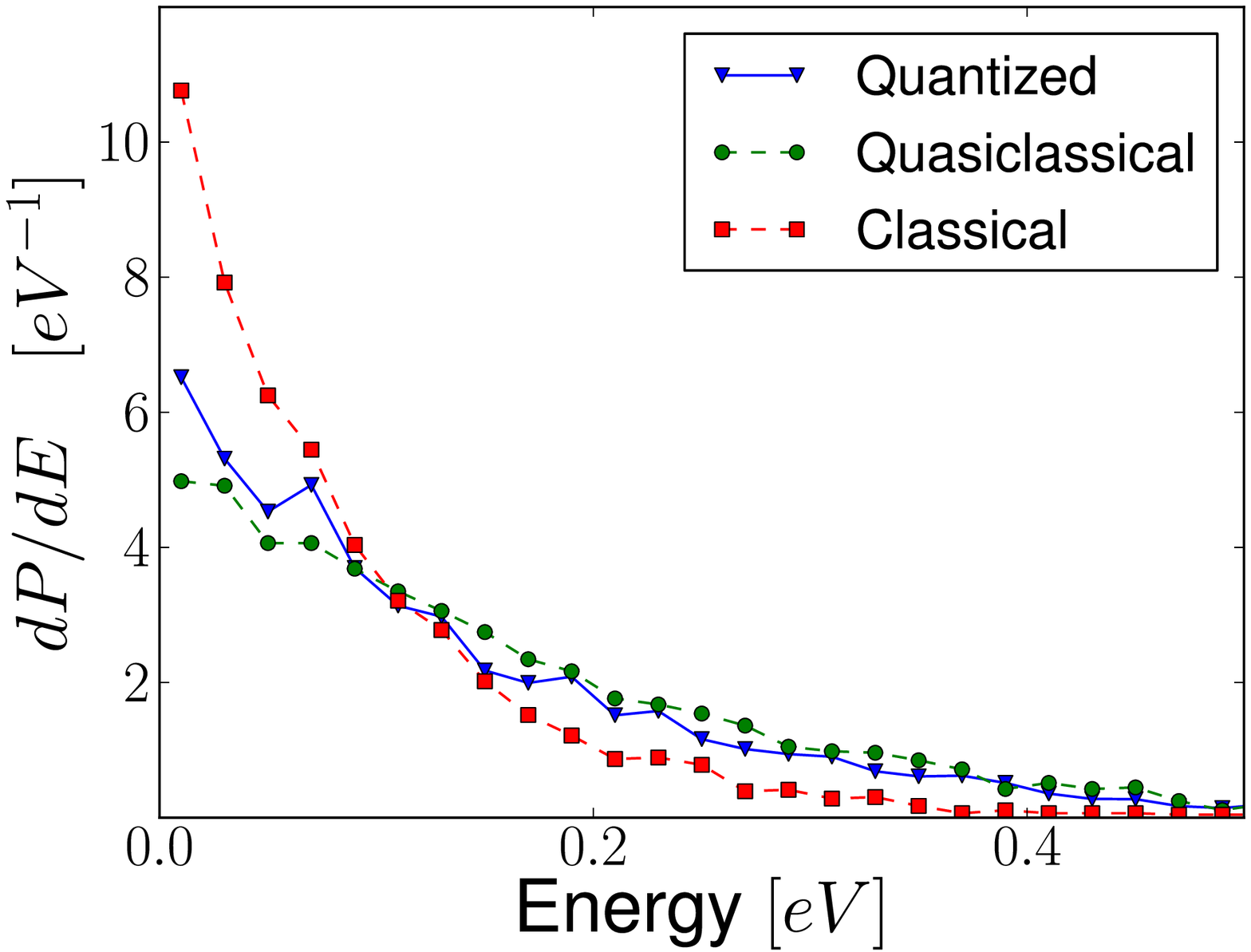}
	  \includegraphics[width=4.25 cm, clip]{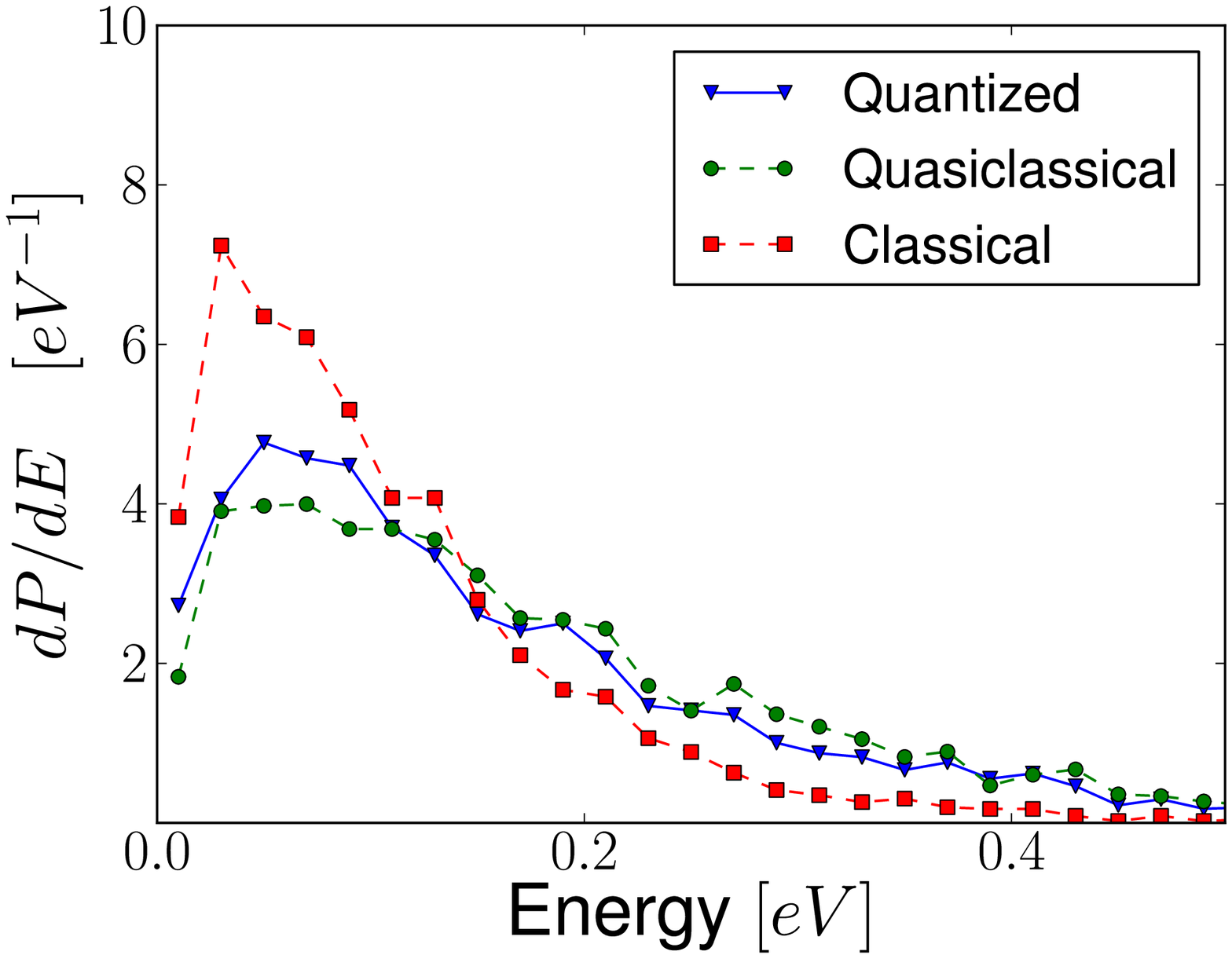}
\caption{The differential probability of desorbed molecule having a given amount of energy as a result of a Gaussian pulse of hot electrons with $T_{max}=4000\;K$ obtained with classical, quasiclassical, and quantized initial conditions. Left is the vibrational energy and right is the total energy.}
\label{fig:dist}
\end{figure}

Although a quantization of the internal mode does not influence the desorption probability it may have a large impact on the distribution of vibrational states of the desorbed molecules. This is illustrated in Fig. \ref{fig:dist} where the distribution of energy is shown for desorbed molecules using the classical, quasiclassical, and quantum initial conditions. Due to the low desorption probabilities we had to start the molecule with a COM momentum of $p=3p_Q$ corresponding to $0.19\;eV$, since otherwise we were not able to get good statistics for the energy distribution of desorbed molecules. However, because of the very weak coupling between the two modes, we do not expect this to have a large influence on the internal energy distribution. The COM energy is not influenced by the initial conditions in the internal mode and the difference in total energy distributions is solely due to differences in the internal mode distributions. It is seen that the quasiclassical initial conditions yields a distribution which is similar to the quantized initial conditions, but with slightly more weight at high lying energies. The classical initial conditions yields a distribution which is inconsistent with a quantized picture, since from Eq. \eqref{P_clas^n} it follows that $dP/dE(E=0)<E_0^{-1}\sim8\;eV^{-1}$.

To see this in more detail we calculate the probabilities of the desorbed molecules being in a particular vibrational state using the method of Sec. \ref{model} and Eq. \eqref{p_n}. The classical initial conditions lead to $p_0>1$ and $p_1<0$ whereas quasiclassical initial conditions give $p_1/p_0=0.22$ and quantized initial conditions give $p_1/p_0=0.092$ which is in agreement with Ref. \onlinecite{struck}. In general, quasiclassical initial conditions tend to overestimate $p_1$ and underestimate $p_0$ and $p_2$ as is seen in Fig. \ref{fig:p_n}. In the present case the error on $p_1/p_0$ is more than a factor of two. For long interaction times and high temperatures the quasiclassical approximation becomes better and we have repeated the above analysis with $T_{max}=6000\;K$, which yields close agreement between the vibrational probabilities resulting from quasiclassical and quantized initial conditions.

\section{Discussion}
In section \ref{model} it was shown that in order to obtain the correct vibrational probabilities for a harmonic oscillator, it is crucial to use quantized initial conditions. However, quasiclassical initial conditions yield good results for the average energy of the harmonic oscillator as well as for the desorption probability of the Morse potential. Naturally, the quasiclassical approximation is highly attractive since it only requires a single initial phase space point, whereas the correctly quantized initial conditions requires a full phase space sampling. In the present work we needed a $6\times6$ grid and $10\times6$ grid of initial phase space points to represent the relevant part of phase space of the harmonic and Morse potentials respectively and quantized initial conditions thus required a factor of 36-60 more calculations than the quasiclassical approach. In general we expect that average quantities are well described by the quasiclassical initial conditions. Similarly, high temperatures (compared to the quantum of oscillation) and long timescales tend to justify the quasiclassical approach.

With CO on Cu(100) as a generic example of a two-dimensional problem with ab initio potentials, we found that quantization of the internal mode had almost no effect on desorption probabilities. However, this is most likely due to the weak coupling between the two modes in the present example, but for reactions with very strong coupling between modes such as associative desorption processes,\cite{luntz1, luntz2} quantization of the internal mode is likely to be important. Furthermore, if one is interested in the final state distribution of vibrational states, it will be crucial to take into account the initial zero point motion of the adsorbate. For example, the fact that hot electron induced associative desorption yields of Hydrogen from Ru(0001) are well described by Langevin dynamics except for too low values of desorbate translational energies,\cite{luntz2, wagner05_etal} may very well be due to initial zero point motion.

It should be mentioned that it is also possible to calculate the friction tensor directly from density functional theory using a basis of Kohn-Sham orbitals.\cite{trail, luntz1, luntz2} While that method is probably more accurate, the present approach based on the reduced density matrix and Newns-Anderson like Hamiltonian Eq. \eqref{H}, gives better access to the physics involved. For example, in the Newns-Anderson framework it is evident that the frictional forces on the center of mass mode and the internal mode have very different physical origins. 
On the other hand, since the Kohn-Sham approach does not make any assumption about the physical nature of the friction, it will automatically include all contributing states and thus give better results when multiple adsorbate states contribute to the friction. The method applied in the present paper only takes into account a single resonance which we assume to have a Lorentzian shape, but the excitation energy is calculated using $\Delta$SCF which gives a much better description than the Kohn-Sham eigenvalues.\cite{gavnholt} At low temperatures, however, the friction is dominated by the projected density of states at the Fermi level which is unlikely to be well described within the wide band limit imposed here.

We have investigated the importance of including the quantized initial state in Langevin dynamics where the friction and stochastic force originate from a thermal bath of hot electrons. In the title we have referred to this as quantum corrected Langevin dynamics, but other quantum corrections may also be important. In particular, for non-quadratic potentials the time evolution is not classical and the Langevin equation should be thought of as a semiclassical approximation to the true dynamics. In principle, the validity of this approximation should always be analyzed in detail for a given potential and time of propagation, but very often one can use a quick 'large $n$' or similar argument to justify the approximation. For example, in the case of CO on Cu(100) we expect the semiclassical approximation to work well, since the Morse potential describing the desorption coordinate has 27 bound states within the $0.57\;eV$ potential well, which gives an energy spacing much smaller than the average adsorbate energy.

Another quantum effect is that of memory in the fluctuating forces. The Markov approximation leading to Eq. \eqref{correlator} completely neglects any correlation between forces at different times and essentially only contains thermal fluctuations. That the Markov approximation has a classical flavor can be seen in the low temperature limit where the fluctuating forces vanish. The Langevin equation with a harmonic potential then gives rise to a decaying average energy: $E(t)=E_0e^{-\eta t/M}$ which is not allowed quantum mechanically, since the average energy can not become less than the zero point energy. This paradox is solved by going beyond the Markov approximation where a small fluctuating force exactly cancels the frictional decay. To get an idea of the range of temperatures where the Markov approximation works, we can estimate the correlation time by $t_c=\hbar/k_BT$. \cite{schmid} The timestep used in the molecular dynamics in this work was $1\;fs$ which corresponds to $T=2900\;K$ and this gives an estimate on the lower temperature limit to the Markov approximation. Memory effects in non-adiabatic dynamics will be explored further in a future paper.

\section{Summary}
We have analyzed the effect of including zero point motion properly in Langevin dynamics with a temperature dependent friction tensor. The method which involves initial phase space sampling, have been compared to a quasiclassical approach where classical initial conditions matching the zero point energy is used. For a harmonic oscillator, the initial conditions is the only quantum mechanical correction since the quantum dynamics becomes classical and we have shown how to obtain vibrational probabilities from the classical energy distribution resulting from Langevin dynamics with phase space sampling. As expected, the result agrees extremely well with an inherently quantum mechanical master equation approach when the initial conditions is included correctly, whereas the quasiclassical approach only tends to a reasonable result after $\sim1\;ps$ of interaction. We have also compared the results of using quantized and 
quasiclassical initial conditions in a Morse potential and found only a little effect on the probability for escaping the potential well. The reason for this can be attributed to the fact that the escape probabilities involves a sum over the continuous set of free states and the quasiclassical approach yields a good description of average quantities.

With CO on Cu(100) as a generic example, we have demonstrated the effect for an adsorbate system with ab initio potentials and electronic friction. The tensor structure of the friction introduces correlation in the fluctuating forces and the nonlinear interaction gives rise to position dependence in the friction. For a model pulse of hot electrons we showed that, compared to the quasiclassical approach, quantized initial conditions both increases the desorption probability and changes the distribution of vibrational states.

\section{Acknowledgments}
This work was supported by the Danish Center for Scientific Computing. The Center for Individual Nanoparticle Functionality (CINF) is sponsored by the Danish National Research Foundation.

\appendix
\section{Path integral representation of the harmonic oscillator density matrix}\label{path}
The derivation of Langevin dynamics is most easily done with a path integral representation of the reduced time dependent density matrix. To see how it works we consider again the harmonic oscillator with a simple degree of freedom and start with the expression \eqref{rho_harmonic}. The two propagators can be written as path integrals resulting in
\begin{align}
\rho(x,y;t)=&\int dx_0dy_0\langle x_0|\rho_0|y_0\rangle\\
&\times\int\mathcal{D}[x(t')]\mathcal{D}[y(t')]e^{iS_0[x(t')]/\hbar-iS_0[y(t')]/\hbar}\notag
\end{align}
with the action
\begin{align}
S_0[x(t')]=\int_0^tdt'\Big(\frac{1}{2}m\dot{x}^2(t')-\frac{1}{2}m\omega^2x^2(t')\Big),
\end{align}
and $x(0)=x_0$, $y(0)=y_0$. Introducing the average path $u(t)=x(t)/2+y(t)/2$ and the fluctuation $v(t)=x(t)-y(t)$ we can do a partial integration on the kinetic term and write the sum of actions
\begin{align}
S_0[x(t')]-S_0[y(t')]&=\int_0^tdt'\Big(m\dot{u}\dot{v}-m\omega^2uv\Big)\notag\\
&=m\dot{u}v\bigg|_0^t-\int_0^tdt'\Big(m\ddot{u}+m\omega^2u\Big)v\notag.
\end{align}
Thus the density matrix becomes
\begin{align}
\rho(x,y;t)=&\int dx_0dy_0\langle x_0|\rho_0|y_0\rangle\\
&\times\int\mathcal{D}[u(t')]e^{im(\dot{u}(t)v(t)-\dot{u}_0v_0)/\hbar}\notag\\
&\times\int\mathcal{D}[v(t')]e^{-iS_0/\hbar}\notag,
\end{align}
where
\begin{align}
S_0[u(t'),v(t')]=-m\int_0^tdt'\Big(\ddot{u}(t')+\omega^2u(t')\Big)v(t')\notag.
\end{align}
It is now straightforward to perform the path integral in $v(t')$ which gives a delta functional on the classical path $\ddot{u}(t')=-\omega^2u(t')$ for the average coordinate.  If we are only interested in the probabilities of finding the particle at a given position we just need the diagonal elements of the density matrix where the end points satisfy $u(t)=x(t)=y(t)$ and $v(t)=0$. In terms of these coordinates the diagonal part of the density matrix becomes
\begin{align}
\rho(u;t)\propto&\int du_0dv_0\langle u_0+v_0/2|\rho_0|u_0-v_0/2\rangle\notag\\
&\times\int\mathcal{D}[u(t')] e^{-im\dot{u}_0v_0/\hbar}\delta(\ddot{u}(t')+\omega^2u(t'))\notag\\
\propto&\int du_0\mathcal{P}(u_0,p_0(u_0,u(t))),
\end{align}
where
\begin{align}
\mathcal{P}(x,p)=\frac{1}{2\pi\hbar}\int dy\langle x+y/2|\rho_0|x-y/2\rangle e^{-ipy/\hbar}
\end{align}
is the Wigner distribution of an initial state described by the density matrix $\rho_0$ and the path integral delta function has been collapsed by noting that for a given $u_0$ there is a unique initial momentum $p_0$ given by
\begin{align}
p_0=m\dot{u}_0=\frac{m\omega}{\sin\omega t}(u(t)-u_0\cos\omega t)
\end{align}
that connects the initial position classically with $u(t)$. The easiest way to determine the normalization is to require that $\int du\rho(u;t)=1$, and the expression is then seen to be identical to Eq. \eqref{density_harmonic}.

%

\end{document}